\documentclass[twoside,fleqn]{article}
\usepackage{amssymb}
\usepackage{npb}
\usepackage{graphics}
\usepackage{graphicx}
\usepackage{epsfig}
\usepackage{psfig}

\newcommand{\be}{\begin{equation}}
\newcommand{\ee}{\end{equation}}

\def\beq{\begin{eqnarray}}    
\def\eeq{\end{eqnarray}}      

\def\be{\beta}

\def\de{\delta}

\def\La{\Lambda}
\def\la{\lambda}
\def\na{\nabla}

\def\si{\sigma}

\def\Ga{\Gamma}

\def\La{\Lambda}

\title{Stability issues in the modified Starobinsky model
}

\author{A.M. Pelinson, \quad I.L. Shapiro,\quad F.I. Takakura
\address{Departamento de Fisica, Universidade Federal de Juiz de
Fora, 36036-330, MG, Brazil}}

\begin{document}

\begin{abstract}
We discuss the stability of the anomaly-induced inflation 
(modified Starobinsky model) with respect to the arbitrary 
choice of initial data and with respect to the small 
perturbations of the conformal factor and tensor modes of 
the metric in the later period of inflation and, partially,
in the present Universe.
\begin{flushright}
\end{flushright}
\end{abstract}

\maketitle


$\,\,\,\,\,\,\,$
The basic principles of the anomaly-induced inflation 
has been explained in \cite{Ilya-OPC} (see also refe\-rences 
and notations therein). 

The advantage of this inflationary model is that it requires 
smaller amount of the cosmological phenomenology than the usual 
inflaton models. To some extent, this model is mainly based on 
the principles of quantum field theory. In particlar, the 
simplest inflationary solution follows 
from the anomaly-induced quantum correction to the
vacuum action $\,{\bar \Ga}[g_{\mu\nu}]$ which is a 
solution of the equation
\beq
\frac{2}{\sqrt{-g}}\,g_{\mu\nu}
\frac{\de}{\de g_{\mu\nu}} {\bar \Ga}\,=\,
(wC^2 + bE + c{\square} R)\,.
\label{main equation}
\eeq

The coefficients $w,b,c$ are defined in \cite{Ilya-OPC},
$\,C^2\,$ is a square of the Weyl tensor and $\,E\,$ is 
an integrand of the Gauss-Bonnet topological term.
This equation admits an explicit solution \cite{rei}.
It is easy to see that this solution contains an ambiguity 
because an arbitrary conformal functional $\,S_c[g_{\mu\nu}]$
plays the role of the ``integration constant'' for the Eq.
(\ref{main equation}). However, this ``integration constant''
does not affect the equation for the conformal factor 
of the metric and in this respect the initial inflationary 
solution follows from the exact effective action. The 
tempered form of expansion which is observed at the later 
inflationary phase is not exact, but it has quite a robust 
background.

The stability of the inflationary solution from the 
initial stage until the graceful exit and the stability of the 
classical solution in the theory with loop corrections represent 
a strong consistency test of the model. Let us start from the 
initial stage of inflation, when the particle content 
$\,N_{0},\,N_{1/2}\,N_1\,$ (number of scalar, fermion and 
vector fields) of the theory provides the stability 
\beq
c>0\,\Rightarrow
\,N_1 \,<\,\frac13\,N_{1/2}\,+\,\frac{1}{18}\,N_{0}\,.
\label{condition}
\eeq
of the exponential solution 
\beq
a(t) \,=\, a_0 \cdot \exp(H_St)\,,\quad
H_S\,=\, \frac{M_P}{\sqrt{-16\pi b}}
\label{flat solution}
\eeq
of Starobinsky \cite{star}. According to \cite{asta}, 
the condition (\ref{condition}) is independent on the
cosmological constant and on the choice of the metric 
$\,k=0\,$ or $\,k=\pm 1$.
For the sake of simplicity we consider $\,k=0\,$ and also 
assume that the cosmological constant is small during 
inflation, such that the last is driven by the quantum 
effects only.

The original Starobinsky model 
deals with the unstable case. The initial data are chosen 
very close to the exponential solution (\ref{flat solution}) 
such that the inflation lasts long enough. Using the 
$0$-$0$ component of the Einstein equations with quantum 
correction, Starobinsky constructed the phase diagram of 
the theory. This phase diagram represent several distinct
attractors, FRW behaviour is one of them and others represent 
physically unacceptable run-away type solutions. In the 
modified version of the model \cite{Ilya-OPC}, the 
inflation starts 
in the stable phase (\ref{condition}). In this case the 
phase diagram, dual to the one of \cite{star}, has the
form shown at the Figure1. This phase portrait indicates
to the unique stable solution (\ref{flat solution}).
Therefore the anomaly-induced 
inflation does not depend on the choice of initial data
and the possible run-away type solutions do not represent 
a threat.
\begin{figure}
\begin{tabular}[c]{clr}%
\mbox{\hspace{0.0cm}} & \resizebox{!}{4.8cm}
{\rotatebox{90}{\rotatebox{90}{\includegraphics[angle=90]{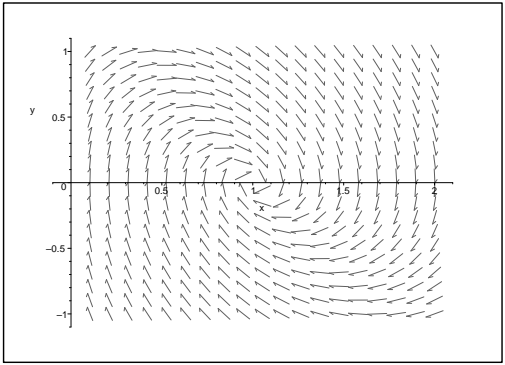}}}} 
\end{tabular}
\caption{\sl The phase diagram for the 
stable version of the Starobinsky model, corresponding to the 
MSSM particle content. All the variables are the same as in 
\cite{star}.}
\end{figure}
Let us now consider another extreme case, that is the 
present Universe dominated by the positive cosmological 
constant $\,\La$. Then the equation of motion 
for the 
conformal factor $\,a(t)\,$ has another approximate 
solution 
\beq
a(t) \,=\, a_0 \cdot \exp(H_St)\,,\quad
H_{c}\,=\,\sqrt{{\Lambda }/{3}}\,.
\label{HH}
\eeq
The present cosmic scale is defined by the  
magnitude of the Hubble parameter $H_0\approx 10^{-42}\,GeV$.
Then the unique active degree of freedom which contributes
to the vacuum quantum effects is a photon, and hence 
$\,(N_{0},\,N_{1/2}\,N_1)=(0,0,1)$, such that the 
inflation is unstable. But this is not all the story, because 
the stability of the low-energy solution (\ref{HH}) does
not follow from the instability of the high-energy 
solution (\ref{flat solution}).
Let us remember that the corresponding equations of motion 
(see, e.g. \cite{asta,Ilya-OPC}) include the fourth derivative 
terms and therefore the stability of the solution (\ref{HH})
with respect to the small perturbations of the conformal 
factor $\,\si(t)=\ln a(t)\,$ can not be seen as a trivial 
occurrence. It may happen that some run-away type solution 
is stable instead. Hence, at this point we meet a very strong 
consistency test for the whole approach. 

Consider the perturbation 
$H \to H + A\,\exp(\la t)$ where $\,\la=const$. 
The solutions of the characteristic
equation for $\la$ have the form \cite{asta}
\beq
\la_1=-4H\,,\qquad
\la_{2/3}=-\frac32\,H\,\pm\,\frac{M_P}{\sqrt{8\pi |c|}}\,i\,.
\label{root 2}
\eeq
It is nice to see that the present-day Universe is not in 
danger due to the higher-derivative terms, because the positive 
cosmological constant provides stability of the solution 
(\ref{HH}). 

The next test is related to the stability 
of the inflationary solution at the last phase, 
when the quantum effects of massive fields temper
the exponential behavior. 
In this case we can use the approximate analytic
method or numerical simulations. The results of both
methods are the same \cite{asta}. Let us briefly 
describe the analytic method. The stability or 
instability with respect to the small perturbations 
depends on the behavior of $\,\si(t)\,$ at the relatively 
small intervals of time,
when the Hubble parameter $\,H\,$ can be treated as a 
constant. Of course, when we move from one such interval 
to another, $\,H\,$ changes providing a source
for the perturbations. The direct calculations give the 
following equation for the perturbations 
$\,\si \to \si + y(\tau)$, where we used ``renormalized'' 
time variable $\,\tau=t/H$, $\,$ $H=const$:
\beq
{\stackrel{....}{y}} + 7\,{\stackrel{...}{y}}
+ 2\Big(6-\frac{b}{c}\Big)\,{\stackrel{..}{y}}
-\frac{8b}{c}\,{\stackrel{.}{y}} 
-\frac{4b}{c}\,\tilde{f}\,y = 0\,
\label{perturbation}
\eeq
At this point we
assumed, as before, a relatively small value of 
the cosmological constant. The last equation has a very 
special form, because all the coefficients are constants
and all but the last have the magnitude 
of the order one. The last coefficient is extremely 
small because of the factor $\,\tilde{f}\ll 10^{-9}$. The 
stability of equation with constant coefficients 
may be explored, e.g. using the Routh-Hurwitz (RH) 
conditions. A priory the RH determinants may have an 
arbitrary sign, but in our case they all turn out to be 
positive, such that the stability of the tempered 
inflation holds until the region $H\approx M_*$. 

Let us consider the stability 
with respect to tensor perturbations of the metric.
In the covariant formalism (see, e.g., \cite{brandenberg})
the evolution of the tensor
degree of freedom is described by the coordinate-dependent 
scalar factor $\,h(t,{\vec r})\,$ of the tensor mode. 
The dynamical equation for $\,h(t)\,$ is very 
complicated \cite{star1,wave}, even for the 
theory of a massless fields. Moreover, this equation
contains an ambiguity due to the conformal functional 
$\,S_c[g_{\mu\nu}]$, which we discussed above. One can 
fix this ambiguity by choosing the proper vacuum for the 
perturbations \cite{wave}. As a result we meet an almost flat 
spectrum of the perturbations, however their 
amplitude may increase very fast. At the same time the
amplification of the amplitudes performs slower 
than the expansion of the conformal factor. As a result
the total metric becomes more and more homogeneous and
isotropic. This is a situation at the initial stage of 
inflation. 

At the last stage, e.g. in the last 65 $\,e$-folds, 
the equation for $\,h(t)\,$ is greatly simplified due
to the enormous number of the total $\,e$-folds between
the beginning and the end of inflation. The typical value 
for $\,\si\,$ depends on the model (see 
\cite{Ilya-OPC}) but it varies between $10^{10}$ and 
$10^{32}$. Obviously, the $\,\si\,$ itself may be 
treated as a very large number in the last 65 inflationary 
$\,e$-folds. This feature greatly simplifies the 
equation for $\,h(t,{\vec r})$. In particular: 
\vskip 1mm

i) The conformal functional $\,S_c[g_{\mu\nu}]$ and the 
choice of the classical action of vacuum have no 
importance. The equation for $\,h(t,{\vec r})$
is completely defined by the universal $\be$-functions
$\,w,b,c\,$ for the vacuum parameters. In particular,
the difference between the equations of \cite{star1}
and \cite{wave} (it is due to the 
different choice of $\,S_c[g_{\mu\nu}]$) disappears in 
this approximation.
\vskip 1mm

ii) The terms with space derivatives 
$\,\na h(t,{\vec r})\,$ are suppressed by the factors of 
$\,\exp (-2\si)\,$ and therefore are negligible \cite{asta}.
\vskip 1mm

iii) The remaining equation for $\,h(t)\,$ has the form
(in the same time variables as in (\ref{perturbation}))
\beq
\stackrel{....}{h} + 6\stackrel{...}{h}
+ 11\stackrel{..}{h} + 6\stackrel{.}{h}
\,-\, \frac{12b}{w\si_f}\,h\,=\,0\,,
\label{wave}
\eeq
where $\,\si_f\,$ is a value of conformal factor 
$\,\si_f\approx\si(t_f)=1/\tilde{f}\,$ corresponding to 
the point of transition from stable to unstable inflation.
Since the magnitude of $\,\si\,$ is huge, one can 
disregard its variation in the last 65 $\,e$-folds.

It is remarkable that the general structure of the 
CEQ. (\ref{wave}) is quite similar to the 
one of the Eq. (\ref{perturbation}) for the 
perturbations of $\,\si(t)$. But it is 
even more remarkable that the solution of 
(\ref{wave}) does not have growing modes. 
One of the roots of the characteristic equation is 
of the order of $\,\tilde{f}\,$ and others of the 
order of one, but all of them have negative real parts.
For the physically reasonable choice of the initial 
data the amplitude of the perturbations almost remains 
constant.
\vskip 1mm

Finally, we can conclude that the stability of the 
inflationary solution with respect to the perturbations
of the conformal factor and the tensor mode of the metric
holds from the initial stage (when the 
quantum fields may be approximately considered massless)
until the scale $M_*$, when the most of the sparticles
decouple and the inflation becomes unstable. 
Indeed, the stability of the model may be jeopardized 
in the transition period, where we expect to 
meet rapid oscillations of the conformal factor which 
should lead to reheating. The main line in the 
further development of the model must be related 
with the quantitative description of the decoupling
and the transition period, investigation of a 
reheating and density perturbations. Indeed, the last 
issue has been already considered \cite{much} in the 
framework of the original Starobinsky model \cite{star}.

\textit{Acknowledgments}: 
Authors are grateful to FAPEMIG for the research grant. 
The work of I.L. Sh. was partially supported by the 
fellowship from CNPq.

\providecommand{\href}[2]{#2}

\end{document}